# The layered iron arsenide oxides $Sr_2CrO_3FeAs$ and $Ba_2ScO_3FeAs$

Marcus Tegel[a], Franziska Hummel[a], Sebastian Lackner[a], Inga Schellenberg[b], Rainer Pöttgen[b], and Dirk Johrendt*[a]

[a] Department Chemie und Biochemie, Ludwig-Maximilians-Universität München, Butenandtstrasse 5–13 (Haus D), 81377 München, Germany

[b] Institut für Anorganische und Analytische Chemie, Westfälische Wilhelms-Universität Münster, Corrensstrasse 30, 48149 Münster, Germany

**Abstract.** The iron arsenide oxides $Sr_2CrO_3FeAs$ and $Ba_2ScO_3FeAs$ were synthesized by high temperature solid state reactions and their crystal structures determined by the X-ray powder diffraction. Their structures are tetragonal ($P4/nmm$; $Sr_2CrO_3FeAs$: $a$ = 391.12(1) pm, $c$ = 1579.05(3) pm; $Ba_2ScO_3FeAs$: $a$ = 412.66(5) pm, $c$ = 1680.0(2) pm, $Z$ = 2) and isotypic to $Sr_2GaO_3CuS$. Iron arsenide layers are sandwiched between perowskite-like metal oxide layers and separated by ~1600 pm, which is much larger compared to the ZrCuSiAs-type '1111' iron arsenide superconductors. The bond lengths and angles within the FeAs layers are adapted to the space requirements of the oxide blocks. Measurements of the magnetic susceptibility, electrical resistivity and temperature-dependent crystal structure show no hint for a structural phase transition or magnetic anomaly in both compounds. $Sr_2CrO_3FeAs$ shows Curie-Weiss paramagnetism above 160 K with an effective magnetic moment of 3.83(3) $\mu_B$ in good agreement with the theoretical value of 3.87 $\mu_B$ for $Cr^{3+}$ ($S$ = 3/2). Antiferromagnetic ordering was detected below $T_N$ ~ 31 K. [57]Fe Mössbauer spectra of $Sr_2CrO_3FeAs$ show one single signal that broadens below the Néel temperature due to a small transferred hyperfine field induced by the magnetic ordering of the chromium atoms. [57]Fe-Mössbauer spectra of $Ba_2ScO_3FeAs$ show single signals which are only subject to weak quadrupole splitting.

*Prof. Dr. Dirk Johrendt
Department Chemie und Biochemie, Universität München
Butenandtstrasse 5–13 (Haus D), D-81377 München, Germany
e-mail: dirk.johrendt@cup.uni-muenchen.de

## Introduction

The discovery of high-$T_c$ superconductivity in iron arsenides [1] has generated an enormous and growing tide of interest [2], which is reflected in about 600 papers already published within one year. Besides outstanding superconducting properties with $T_c$'s up to 55 K [3] and very high critical fields of at least 70 T [4], it is certainly also the richness of crystal chemical and physical properties, that has pushed this class of materials in the spotlight of interest.

Up to now, the family of iron-based superconductors consists of rather simple ternary or quaternary compounds with long known crystal structures, namely LaOFeAs with the ZrCuSiAs-type structure [1, 5], $BaFe_2As_2$ with $ThCr_2Si_2$-type structure [6-8], LiFeAs with PbFCl-type structure [9, 10] and β-FeSe with the anti-PbO structure [11, 12]. The currently by far most investigated compounds are the $RE$FeAsO ('1111', $RE$ = rare earth) and $A$Fe$_2$As$_2$ ('122', $A$ = alkaline earth) based systems.

Superconductivity emerges in the FeAs layers built up by edge-sharing FeAs$_{4/4}$ tetrahedra, and is assumed to be unconventional [13, 14] because of the relatively high critical temperatures and the proximity to structural and magnetic transitions. The latter ones only occur in the non-superconducting 'parent compounds' like LaFeAsO [15] or BaFe$_2$As$_2$ [6] and have to be at least partially suppressed by doping or by external pressure before superconductivity arises [1, 8]. This scenario is reminiscent of the high-$T_c$ cuprates, where also an antiferromagnetic ground-state of a layered parent compound like La$_2$CuO$_4$ becomes unstable upon doping before superconductivity appears [16]. However, this analogy is limited by the fact, that the doping of a Mott-insulator as in the case of the cuprates is of a very different nature compared to doping the metallic iron arsenides. Indeed, the metallic property of the parent compounds and the multi-band character of the Fermi-surfaces constitute significant differences between the cuprate and iron arsenide superconductors. It has also been shown that the anisotropy of the iron arsenide superconductors is significantly lower compared to the cuprates, which have even higher $T_c$'s [17].

Even though one should be careful in transferring principles from the cuprates, it is an important task to search for new iron arsenides with lower dimensional structures, i. e. with larger FeAs interlayer distances. Several pnictide oxide structures are considerable candidates [18], but also materials derived form copper sulfide oxides with isoelectronic CuS layers [19, 20] are very promising. The first reported compound was $Sr_3Sc_2O_5Fe_2As_2$ [21] with the known structure of $Sr_3Fe_2O_5Cu_2S_2$ [20]. The iron arsenide is not superconducting and shows no structural anomaly or magnetic ordering. Then superconductivity at 17 K [22] has been found in the iron phosphide $Sr_2ScO_3FeP$ with the $Sr_2GaO_3CuS$-type [19] structure. This $T_c$ is considerably higher in comparison with the ZrCuSiAs-Type phosphide oxides like LaFePO (4-7K) and may promise even higher $T$c's in analogue arsenides.

In this paper, we report on the iron arsenide oxides $Sr_2CrO_3FeAs$ and $Ba_2ScO_3FeAs$, both sharing the key feature of tetragonal iron arsenide layers with other iron arsenide superconductors, but the interlayer distances are much larger due to separation by perowskite-like metal oxide blocks. Their crystal structures were determined by X-ray powder diffraction and the physical properties were characterized by magnetic susceptibility and electrical resistivity measurements and $^{57}Fe$ Mössbauer spectroscopy experiments. The electronic structure of the new compound $Ba_2ScO_3FeAs$ is compared with the parent compounds of the superconducting iron arsenides.

## Results

*Crystal structure*

The crystal structures of $Sr_2CrO_3FeAs$ and $Ba_2ScO_3FeAs$ are isotypic to the oxide sulfide $Sr_2GaO_3CuS$ [19] in the space group $P4/nmm$. Figure 1 shows X-ray powder patterns, which could be completely fitted with one single phase in the case of $Sr_2CrO_3FeAs$ using starting parameters from ref. [19]. Minor impurities of $Sc_2O_3$ and FeAs were detected in the $Ba_2ScO_3FeAs$ sample. Crystallographic data, selected bond lengths and parameters of the Rietveld fits are compiled in Table 1. Further details of the structure determinations may be obtained from: Fachinformationszentrum Karlsruhe, D-76344 Eggenstein-Leopoldshafen (Germany) by quoting the Registry No's CSD-###### ($Sr_2CrO_3FeAs$) and CSD-###### ($Ba_2ScO_3FeAs$).

The crystal structure of $Sr_2CrO_3FeAs$ and $Ba_2ScO_3FeAs$ is shown in Figure 2. The FeAs layers perpendicular to the *c*-axis are separated by strontium atoms from chromium oxide and strontium oxide layers according to the stacking order $(CrO_2)(SrO)(SrO)(CrO_2)$. The Fe–As bonds are shorter and the $FeAs_{4/4}$ tetrahedra are less distorted in $Sr_2CrO_3FeAs$ in comparison with $Ba_2ScO_3FeAs$. Also the Fe–Fe distance is significantly shorter in the strontium compound (276.6 pm) than in the barium compound (291.8 pm). This may be attributed to the larger required space of the barium ions and to a smaller extent the scandium ions, and it shows that the geometry of the FeAs layer is flexible and can be adapted to the oxide blocks. The interlayer distance of the iron arsenide layers equals the $c$ lattice parameter for this structure type, thus $Ba_2ScO_3FeAs$ ($c = 1680$ pm) exhibits the highest iron arsenide interlayer distance so far. The powder patterns recorded at lower temperatures revealed no structural phase transitions for both compounds down to 10 K. At 10 K, the corresponding $a$ lattice parameter is decreased by 0.73 pm (0.92 pm) and the $c$ lattice parameter by 8.4 pm (6.0 pm) for $Sr_2CrO_3FeAs$ ($Ba_2ScO_3FeAs$).

*Magnetism and resistivity*

Figure 3 shows the magnetic susceptibility of $Sr_2CrO_3FeAs$ between 1.8 and 360 K. The compound obeys the Curie-Weiss law above ~ 160 K. An effective magnetic moment of 3.83(3) $\mu_B$ was found, which is in good agreement with the theoretical spin-only value of 3.87 $\mu_B$ for $Cr^{3+}$ ($S = 3/2$). The paramagnetic temperature is −141(3) K. Antiferromagnetic ordering is discernible, the highest susceptibility was reached at $T_N$ ~ 31 K. The origin of the anomaly around 120 K is not yet clear, but probably due to traces of ferromagnetic contamination. $Ba_2ScO_3FeAs$ is Pauli-paramagnetic in the temperature range between 1.8 and 300 K (not shown). Neither compound shows any sign of anomaly in the course of the magnetic susceptibility at any applied external magnetic field. The resistivity of both $Sr_2CrO_3FeAs$ and $Ba_2ScO_3FeAs$ is depicted in Figure 4. Both compounds are poor metals over the whole measured temperature range and show no anomaly. Both magnetism and resistivity therefore indicate that there is no occurrence of a spin-density-wave anomaly in any of the compounds.

*Mößbauer spectroscopy*

The $^{57}$Fe Mössbauer spectra collected for the $Sr_2CrO_3FeAs$ sample at various temperatures are shown in Figure 5 together with transmission integral fits. The corresponding fitting parameters are summarized in Table 2. As expected from the crystal structure (one single Fe site), the spectra at 298, 77, and 40 K are well reproduced with single signals, which are subject to weak quadrupole splitting due to the non-cubic site symmetry ($\overline{4}m2$) of the iron atoms. Within the standard deviations, the observed isomer shift, the experimental line width, and the quadrupole splitting of $Sr_2CrO_3FeAs$ are almost identical with those obtained for $Sr_3Sc_2O_5Fe_2As_2$ [23], indicating a similar electronic situation for the iron atoms in both arsenide oxides. The observed isomer shift is in line with the recently reported [57]Fe data for LaFePO [24], $SrFe_2As_2$ [25], and $BaFe_2As_2$ [26]. The increase of the isomer shift with decreasing temperature (0.29 → 0.45 mm/s) is due to a second order Doppler shift (SODS), well known for iron compounds.

In contrast to $Sr_3Sc_2O_5Fe_2As_2$ [21], where no magnetic ordering is observed down to 4.2 K, the chromium atoms in $Sr_2CrO_3FeAs$ reveal antiferromagnetic ordering. This is also reflected in the [57]Fe Mössbauer spectra at 20 and 4.2 K, i. e. well below the Néel temperature. In the magnetically ordered regime we observe significant broadening of the Mössbauer signal and the fits revealed increased line width and quadrupole splitting parameters. Since the transferred field ($B_{hf}$) is small, independent refinement of all parameters, $\delta$, $\Gamma$, $\Delta E_Q$, and $B_{hf}$ showed strong correlations. In order to get an estimate for the transferred field, the 4.2 K spectrum was fitted with various fixed values for the hyperfine field. A reasonable fit was obtained for a fixed hyperfine field of 0.5 T and the refined values $\delta$ = 0.45(1) mm/s, $\Gamma$ = 0.84(2) mm/s, and $\Delta E_Q$ = 0.24(1) mm/s. From these fitting tests we estimate a transferred hyperfine field of 0.5±0.2 T at 4.2 K.

Figure 6 shows the [57]Fe Mössbauer spectra of the $Ba_2ScO_3FeAs$ sample at various temperatures together with transmission integral fits. The fitting parameters are listed in Table 2. In agreement with the single Fe site in the crystal structure, the spectra at 298, 77, and 4.2 K show single signals which are subject to weak quadrupole splitting. The 77 K isomer shifts of $Ba_2ScO_3FeAs$ (Table 2), $Sr_3Sc_2O_5Fe_2As_2$ [23], and $Sr_2CrO_3FeAs$ are almost identical. We can thus assume a similar electronic situation within the tetrahedral [$Fe_2As_2$] layers of the three structures. They are also comparable to LaFePO [24] and $SrFe_2As_2$ [25]. The increase of the isomer shift with decreasing temperature (0.35 → 0.50 mm/s) results from a second order Doppler shift. The spectra give no hint for magnetic ordering of the iron moments down to 4.2 K. $Ba_2ScO_3FeAs$ and $Sr_3Sc_2O_5Fe_2As_2$ show a lower quadrupole splitting parameter (0.20 mm/s for both compounds) than $Sr_2CrO_3FeAs$ (0.42 mm/s) at 4.2 K. Even smaller values of $\Delta E_Q$ = 0.19 and -0.04 mm/s occur for LaFePO and $BaFe_2As_2$, respectively.

*Electronic structure*

The electronic structure of the iron arsenide superconductors have frequently been described [27-31] and some (maybe hidden) electronic order with respect to superconductivity is still under discussion [32]. Also $Sr_3Sc_2O_5Fe_2As_2$ and the superconducting phosphide $Sr_2ScO_3FeP$ have been studied recently [33,34]. In all cases, the calculations predict a Fermi surface consisting of hole-like pockets at the center of the Brillouin zone ($\Gamma$) and similar electron-like pockets at the corners (M), which become largely congruent by a vector ($\pi,\pi$). This Fermi surface nesting is an unstable situation and typically favors charge- or spin-density ordering, often connected with structural distortions. We have calculated the band structure of $Ba_2ScO_3FeAs$ in order to see whether the rather large metal oxide building blocks between the FeAs layers induces any appreciable perturbance of the bands close to the Fermi level and thus the Fermi surface. The band structure of $Ba_2ScO_3FeAs$ (Figure 7, top) shows the typical features known from the iron arsenide superconductors [27]. Iron $3d$ band contributions (emphasized in Figure 7) clearly dominate the vicinity of the Fermi level, where no significant contributions of the oxide layers are found. Also the Fermi surface (Figure 7, bottom), reveals the typical cylinders around $\Gamma$ and M and provides the unstable nesting situation as mentioned above. Thus from the view of the electronic structure, one would expect structural and magnetic anomalies in $Ba_2ScO_3FeAs$. The origin of the absence of any structural or magnetic instability and superconductivity in this and the related compounds is not clear and currently under investigation.

## Conclusion

The iron arsenide oxides $Sr_2CrO_3FeAs$ and $Ba_2ScO_3FeAs$ crystallize in the tetragonal $Sr_2GaO_3CuS$-type structure (space group $P4/nmm$) and exhibit the largest distance between the FeAs layers so far due to separation by large perowskite-like metal oxide blocks. The chromium compound shows paramagnetism in agreement with $Cr^{3+}$ ($S =$ 3/2) and antiferromagnetic ordering below 31 K, whereas $Ba_2ScO_3FeAs$ is Pauli-paramagnetic. Both compounds are poor metals and show no structural or magnetic anomalies, even though the electronic structure predicts a nested Fermi surface. [57]Fe-Mössbauer spectra are in agreement with the absence of magnetic ordering at the iron site. In the case of $Sr_2CrO_3FeAs$, a small hyperfine field is transferred to the iron nuclei from the adjacent magnetically ordered $CrO_2$ layers.

**Note added:** While we were finalizing our manuscript, we recognized an unpublished manuscript by H. Ogino et al. [35]. Therein, the iron arsenide oxides $Sr_2MO_3FeAs$ ($M$ = Sc, Cr) are reported. The resistivity and susceptibility data are slightly different from our results.

## Acknowledgments

We thank Dipl.-Chem. Bele Boeddinghaus and Prof. Thomas Fässler for help with the magnetic measurements. This work was financially supported by the Deutsche Forschungsgemeinschaft.

## Experimental

*Synthesis*

$Sr_2CrO_3FeAs$ and $Ba_2ScO_3FeAs$ were synthesized by heating stoichiometric mixtures of strontium, chromium, iron (III) oxide, arsenic oxide and barium, scandium, iron (III) oxide and arsenic oxide, respectively in alumina crucibles sealed in silica ampoules under an atmosphere of purified argon. The mixtures were heated to 1323 K at a rate of 80 K/h, kept at this temperature for 60 h and cooled down to room temperature. The products were homogenized in an agate mortar, pressed into pellets and sintered at 1323 K for 60 h, reground, pressed into pellets and sintered again at 1323 K for 50 h.

*Crystal structure determination*

Powder patterns were recorded on STOE STADI P Debye-Scherrer X-Ray diffractometers (Mo-$K_{\alpha 1}$ and Cu-$K_{\alpha 1}$ radiation, respectively, Ge-111 monochromator, Debye-Scherrer or transmission geometry). Rietveld refinements of both samples were performed with the TOPAS package [36] using the fundamental parameters approach as reflection profiles (convolution of appropriate source emission profiles with axial instrument contributions as well as crystallite microstructure effects). In order to describe small peak half width and shape anisotropy effects, the approach of *Le Bail* and *Jouanneaux* [37] was implemented into the TOPAS program and the according parameters were allowed to refine freely. Preferred orientation of the crystallites was described with a spherical harmonics function of $4^{th}$ or $8^{th}$ order. In order to check for structural phase transitions, powder patterns between 10 and 300 K were recorded on a Huber G670 Guinier imaging plate diffractometer (Cu-K$_{\alpha1}$ radiation, Ge-111 monochromator), equipped with a closed-cycle He-cryostat. These patterns were also refined with the TOPAS package using the approach described in Ref. [26].

*Resistivity measurements*

Sintered powder pellets (diameter = 6 mm, height ≈ 1.4 mm) of the samples were contacted with copper wires and silver conducting paint, and the electrical resistivity was measured between 10 and 320 K using a dc four-point current-reversal method [38].

*Magnetic measurements*

The magnetic properties were studied using a commercial SQUID magnetometer (Quantum Design MPMS-5) with external magnetic fields up to 50 kOe. Susceptibilities were measured at 100, 500, 1000, 5000 and 10000 Oe and hysteresis loops were taken at various temperatures to check for ferromagnetic ordering or impurity phases.

*$^{57}$Fe Mössbauer spectroscopy*

A $^{57}$Co/Rh source was available for the $^{57}$Fe Mössbauer spectroscopy investigations. The quoted values of the isomer shifts are given relative to this material. The $Sr_3Sc_2O_5Fe_2As_2$ sample was placed in a thin-walled PVC container. The measurements were conducted in the usual transmission geometry at temperatures between 4.2 and 298 K. The source was kept at room temperature. The total counting time was 1 day per spectrum. Fitting of the spectra was carried out with the NORMOS-90 program system [39].

*Computational details*

Electronic structure calculations for $Ba_2ScO_3FeAs$ were performed with the *WIEN2k* program package [40] using density functional theory within the full-potential *LAPW* method and the local density approximation (*LDA*). Detailed descriptions for both methods are given elsewhere [41]. Full-potential *LAPW*, which was used for the *WIEN2k* calculations, is based on the muffin-tin construction with non-overlapping spheres. In the interstitial region between spheres, the potential is represented by a plane wave expansion. Because of the great flexibility and accuracy of this expansion for the potential and charge density, a very high numerical accuracy is achieved for the *LAPW* method. Mixed *LAPW* and *APW + lo* (*lo* = local orbitals) basis sets were used to increase the efficiency of the *APW* linearization [42]. Further technical details can be found in ref. [43] and the monograph of Singh [44]. The total energies and charge densities of the SCF cycles converged to changes smaller than $1 \times 10^{-4}$ Ryd cell$^{-1}$ and 4374 *k*-points ($27 \times 27 \times 6$ mesh) were used in the irreducible wedges of the Brillouin zones. The basis sets consisted of 4192 plane waves up to a cutoff $R_{mt}K_{max} = 8.0$. The atomic sphere radii $R_{mt}$ were 2.50 (Ba), 1.97 (Sc), 2.43 (Fe), 2.16 (As) and 1.75 au (O). Fermi surfaces were visualized with XCrysDen [45].

# Tables

## Table 1. Crystallographic data for Sr$_2$CrO$_3$FeAs and Ba$_2$ScO$_3$FeAs at 297 K.

| compound | Sr$_2$CrO$_3$FeAs | Ba$_2$ScO$_3$FeAs |
|---|---|---|
| space group | $P4/nmm$ ($o2$) | $P4/nmm$ ($o2$) |
| molar mass (g mol$^{-1}$) | 406.001 | 498.375 |
| lattice parameters (pm) | $a = 391.12(1)$ | $a = 412.66(5)$ |
| | $c = 1579.05(3)$ | $c = 1680.0(2)$ |
| cell volume, (nm$^3$) | 0.24156(1) | 0.28608(7) |
| density, (g cm$^{-3}$) | 5.582(1) | 5.786(1) |
| $\mu$ (mm$^{-1}$) | 76.869 | 25.109 |
| $Z$ | 2 | 2 |
| data points | 9600 | 5350 |
| reflections (main phase) | 108 | 259 |
| $d$ range | 1.003 – 15.790 | 0.7631 – 16.800 |
| atomic variables | 13 | 12 (1 constraint) |
| profile variables | 5 | 5 |
| anisotropy variables | 12 | 24 |
| background variables | 48 | 48 |
| variables of impurity phases | 0 | 20 |
| other variables | 8 | 12 |
| R$_P$, $w$R$_P$ | 0.014, 0.018 | 0.036, 0.047 |
| R$_{bragg}$, $\chi^2$ | 0.002, 0.969 | 0.008, 1.186 |
| wght. Durbin-Watson $d$ stat. | 1.707 | 1.140 |

**Atomic parameters:**

| | | |
|---|---|---|
| Sr1/Ba1 | $2c$ (¾, ¾, $z$); $z = 0.1957(2)$; $U_{iso} = 83(5)$ | $2c$ (¾, ¾, $z$); $z = 0.1860(1)$; $U_{iso} = 163(6)$ |
| Sr2/Ba2 | $2c$ (¾, ¾, $z$); $z = 0.4150(1)$; $U_{iso} = 98(7)$ | $2c$ (¾, ¾, $z$); $z = 0.4145(1)$; $U_{iso} = 129(6)$ |
| Cr1/Sc1 | $2c$ (¼, ¼, $z$); $z = 0.3108(2)$; $U_{iso} = 88(9)$ | $2c$ (¼, ¼, $z$); $z = 0.3092(4)$; $U_{iso} = 61(14)$ |
| Fe1 | $2a$ (¼, ¾, 0); $U_{iso} = 77(8)$ | $2a$ (¼, ¾, 0); $U_{iso} = 190(11)$ |
| As1 | $2c$ (¼, ¼, $z$); $z = 0.0899(2)$; $U_{iso} = 79(8)$ | $2c$ (¼, ¼, $z$); $z = 0.0777(2)$; $U_{iso} = 103(11)$ |
| O1 | $4f$ (¼, ¾, $z$); $z = 0.2931(4)$; $U_{iso} = 78(22)$ | $4f$ (¼, ¾, $z$); $z = 0.2898(7)$; $U_{iso} = 262(32)$ |
| O2 | $2c$ (¼, ¼, $z$); $z = 0.4313(6)$; $U_{iso} = 88(30)$ | $2c$ (¼, ¼, $z$); $z = 0.427(1)$; $U_{iso} = 262(32)$ |

**Selected bond lengths (pm) and angles (degrees):**

| | | |
|---|---|---|
| Sr/Ba – O | 242.7(9)×1; 248.8(4)×4; | 266(2)×1; 270.2(8)×4; |
| | 274.4(5)×4; 277.8(1)×4 | 292.6(2)×4; 294.0(9)×4 |
| Cr/Sc – O | 190(1)×1; 197.5(1)×4 | 198(2)×1; 208.9(2)×4 |
| Fe – Fe | 276.6(1)×4 | 291.8(1)×4 |
| Fe – As | 241.7(2)×4 | 244.1(2)×4 |
| As – Fe – As | 110.2(1)×4; 108.0(1)×2 | 106.6(1)×4; 115.4(1)×2 |
| O – Cr/Sc – O | 88.9(1)×4; 98.1(2)×4; 163.7(4)×2 | 88.6(1)×4; 99.0(4)×4; 162.1(7)×2 |

Table 2. Fitting parameters of $^{57}$Fe Mössbauer spectra at different temperatures. $\Gamma$: experimental line width, $\delta$: isomer shift; $\Delta E_{Q}$: electric quadrupole splitting parameter (for details see text).

**Sr$_2$CrO$_3$FeAs**

| T / K | $\Gamma$ / mms$^{-1}$ | $\delta$ / mms$^{-1}$ | $\Delta E_{Q}$ / mms$^{-1}$ |
|---|---|---|---|
| 298 | 0.37(1) | 0.29(1) | 0.22(1) |
| 77 | 0.36(1) | 0.42(1) | 0.24(1) |
| 40 | 0.31(1) | 0.43(1) | 0.23(1) |
| 20 | 0.48(1) | 0.43(1) | 0.28(1) |
| 4.2 | 0.67(2) | 0.45(1) | 0.43(1) |

**Ba$_2$ScO$_3$FeAs**

| T / K | $\Gamma$ / mms$^{-1}$ | $\delta$ / mms$^{-1}$ | $\Delta E_{Q}$ / mms$^{-1}$ |
|---|---|---|---|
| 298 | 0.39(1) | 0.35(1) | 0.21(1) |
| 77 | 0.45(1) | 0.47(1) | 0.20(1) |
| 4.2 | 0.43(2) | 0.50(1) | 0.20(1) |

**Figure Captions**

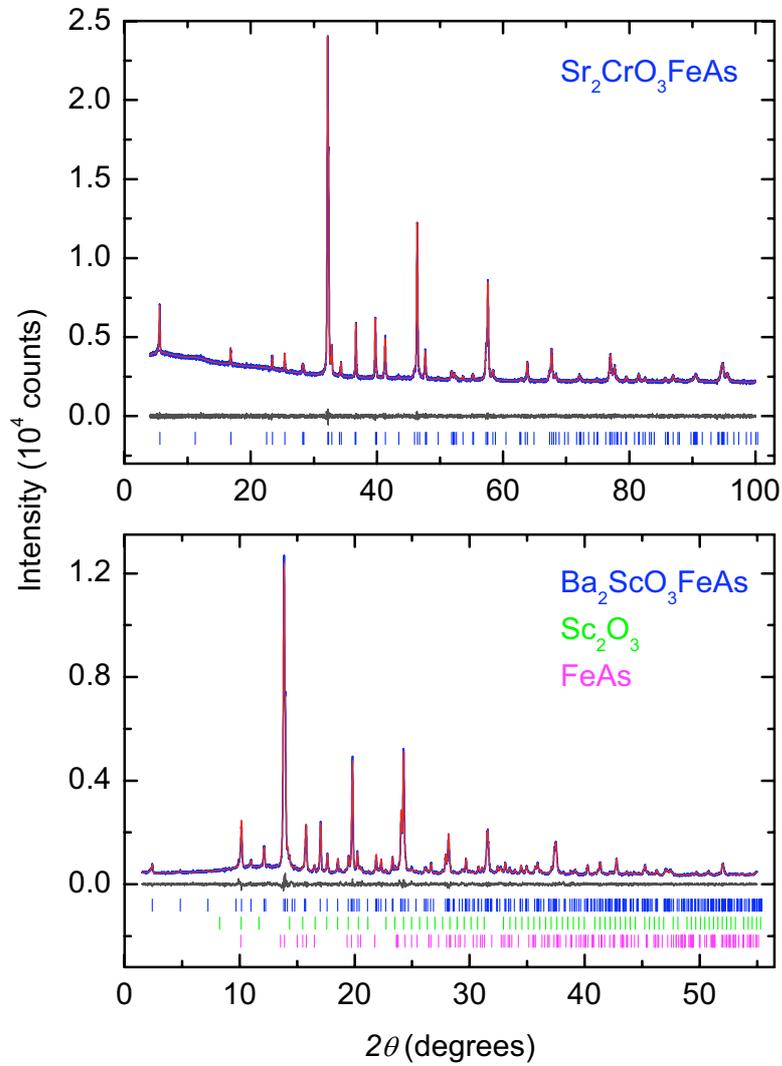

**Fig. 1.** Powder pattern (blue) and Rietveld fit (red) of $Sr_2CrO_3FeAs$ and $Ba_2ScO_3FeAs$ at 297 K (space group $P4/nmm$). Bottom: $Sc_2O_3$ and FeAs were included in the refinement as minor impurity phases.

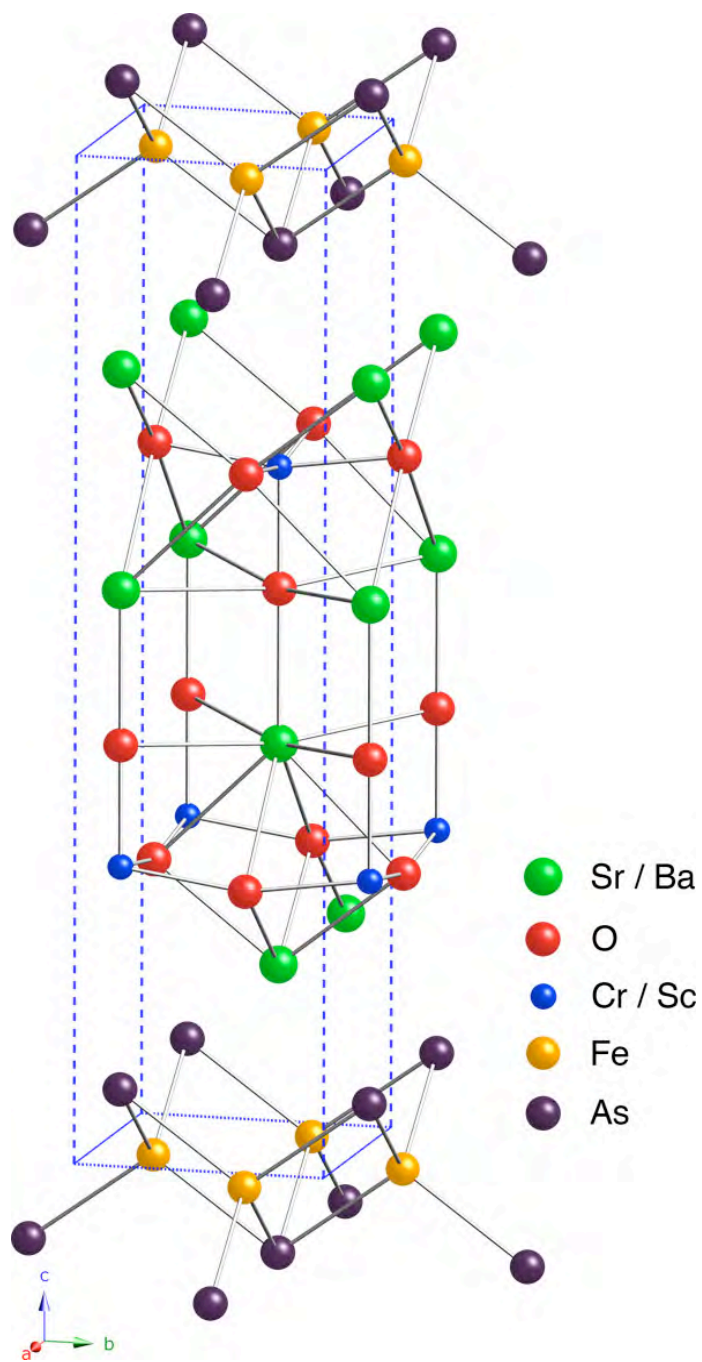

**Fig. 2**. The crystal structure of Sr$_2$CrO$_3$FeAs and Ba$_2$ScO$_3$FeAs.

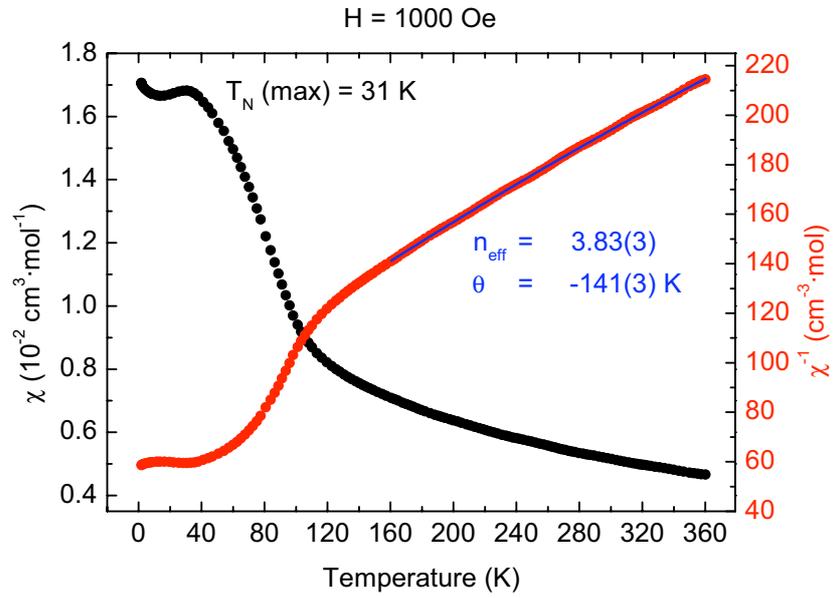

**Fig. 3.** Magnetic susceptibility (black), inverse susceptibility (red) and Curie-Weiss fit (blue) of $Sr_2CrO_3FeAs$.

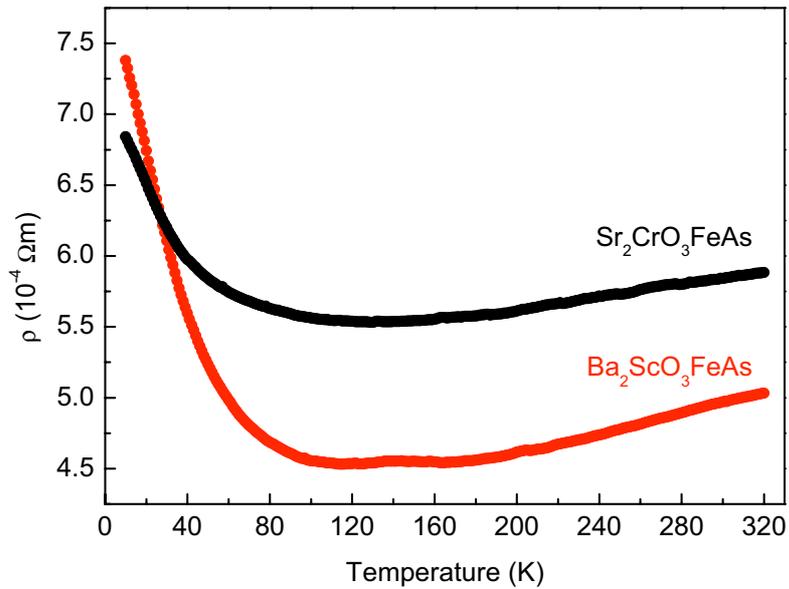

**Fig. 4.** Resistivity of $Sr_2CrO_3FeAs$ (black) and $Ba_2ScO_3FeAs$ (red).

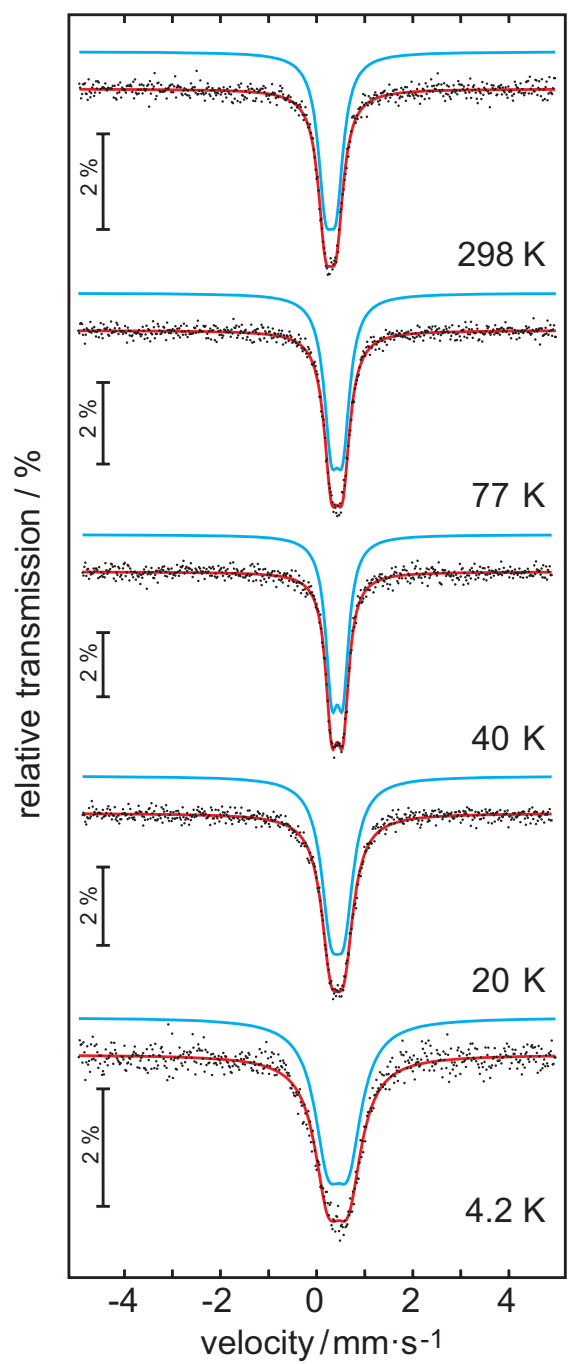

**Fig. 5.** Experimental and simulated $^{57}$Fe Mössbauer spectra of $Sr_2CrO_3FeAs$ at various temperatures.

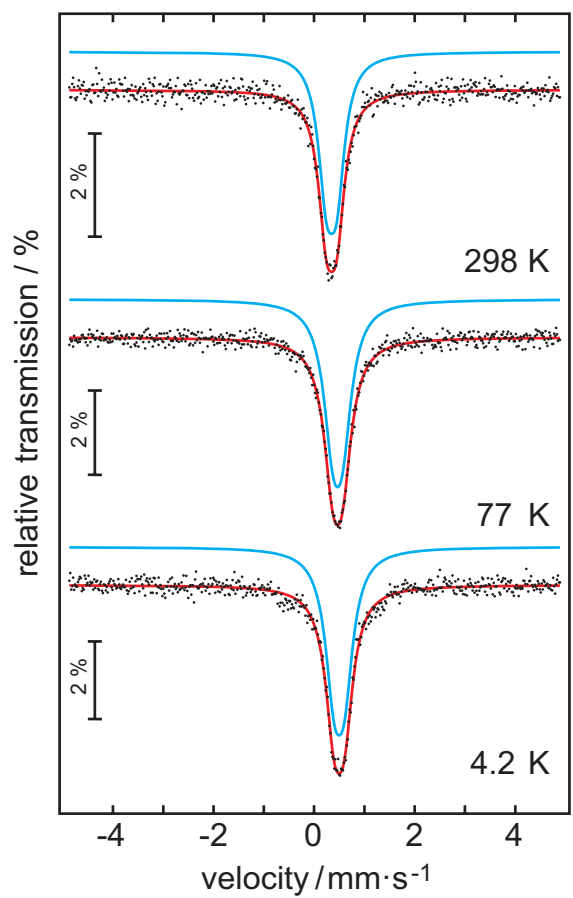

**Fig. 6.** Experimental and simulated $^{57}$Fe Mössbauer spectra of Ba$_2$ScO$_3$FeAs at various temperatures.

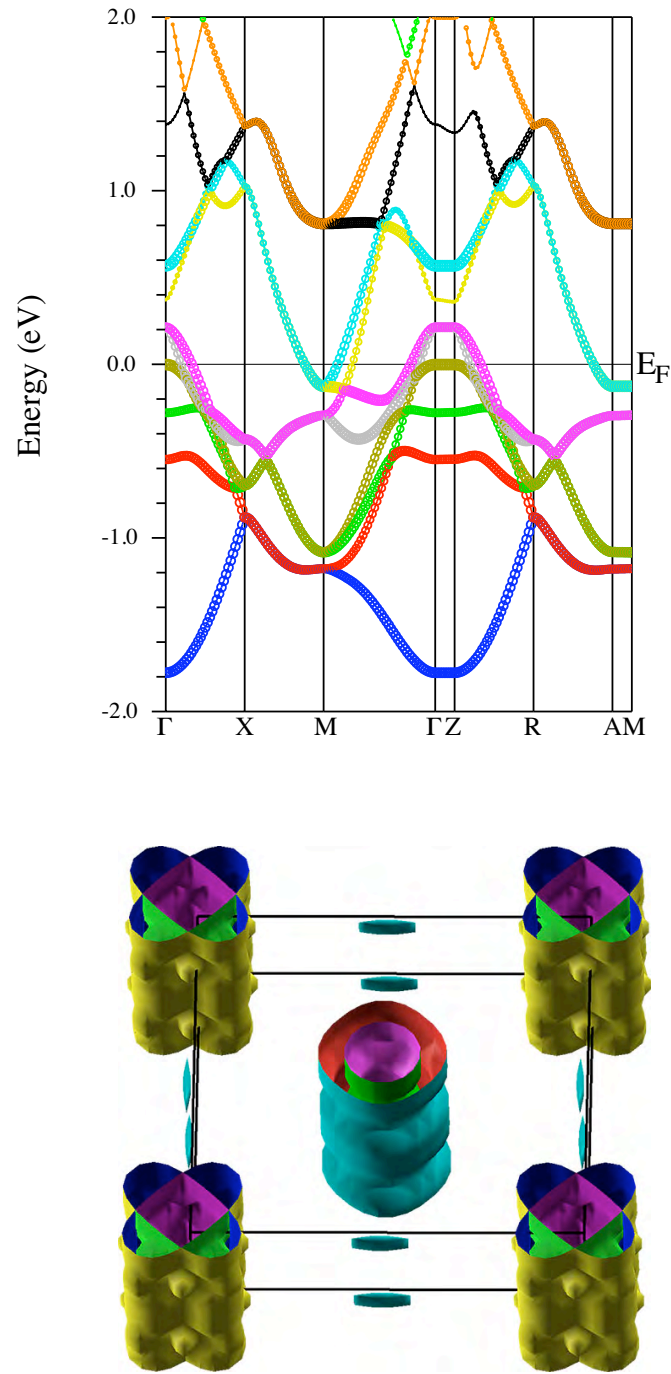

Figure 7. Top: Band structure of Ba$_2$ScO$_3$FeAs with the iron 3$d$ contributions emphasized. Bottom: The Fermi surface showing the typical cylinders around the center ($\Gamma$, hole-like) and the corners ($M$, electron-like).